\documentclass[jap,graphicx, reprint]{revtex4-1} 
\usepackage{amsfonts, amsmath, graphicx}
\usepackage[version=3]{mhchem}

\begin{document} 
\title{Dirac semimetal thin films in in-plane magnetic fields} 

\author{Zhuo Bin Siu}
\affiliation{Computational Nanoelectronics and Nanodevices Laboratory, Electrical and Computer Engineering Department, National University of Singapore, Singapore} 
\author{Mansoor B. A. Jalil} 
\affiliation{Computational Nanoelectronics and Nanodevices Laboratory, Electrical and Computer Engineering Department, National University of Singapore, Singapore} 
\author{Seng Ghee Tan} 
\affiliation{Data Storage Institute, Agency for Science, Technology and Research (A*STAR), Singapore} 

\begin{abstract}
In this work, we study the effects of in-plane magnetic fields on thin films of the Dirac Semimetal (DSM) \ce{Na3Bi} where one of the in-plane directions is perpendicular to the $k$-separation between the two Weyl nodes that exist for each spin orientation. We show numerically that the states localized near the surfaces of these thin films are related to the Fermi arc states in semi-infinite slabs. Due to the anisotropy between the two in-plane directions, the application of a magnetic field along these directions have differing effects.  A field parallel to the $k$ space separation between the Weyl nodes leads to a broadening of the surface state band and the formation of an energy plateau, while a perpendicular field shifts the energy where the hole and particle bands meet upwards, and sharpens the tips of the bands. We illustrate the effects of these changes to the dispersion relation by studying the transmission from a source segment without a magnetic field, to a drain segment with an in-plane magnetic field for various combinations of field and source-drain interface directions.      
\end{abstract} 

\maketitle

\section{Introduction} 

The Dirac Semimetal (DSM) \cite{PRB85_195320, PRB83_205101, PRL108_140405, PRB88_125427} is a recently discovered topologically non-trivial state  which has attracted much attention.  Like the surface states of the more established three-dimensional topological insulators (TI) \cite{RMP82_3045, JPSJ82_102001} , the low energy spectrum of DSMs take the form of Dirac cones. Unlike the Dirac cones in three-dimensional TI surface states which have linear dispersion only in the two dimensions in the plane of a TI \textit{surface}, the DSM Dirac cones disperse linearly in all three $k$-space dimensions in \textit{bulk} DSMs about the Weyl nodes. Moreover, unlike in conventional TIs where there is a odd number of Dirac points, the Weyl nodes in DSMs occur in pairs with one member of each pair acting as a Berry curvature source and the other member acting as a Berry curvature sink.   These pairs of Weyl nodes in DSMs give rise to surface states in the form of Fermi arcs linking the two points in $k$-space when the bulk DSM is truncated perpendicular to the direction of the $k$-space separation between the Weyl nodes.  These Weyl nodes are topologically stable against perturbations which preserve the translational symmetry. To date, two materials, \ce{Cd3As2} \cite{NatMat13_677, PRL113_027603, NatComm5_3786, NatMat13_851} and \ce{Na3Bi} \cite{Sci343_864, APL105_031901, Sci347_294}  have been experimentally confirmed to host the DSM state. 

The effects of electric fields applied perpendicular to the surfaces of quasi-two dimensional thin films and quasi-one dimensional \ce{Na3Bi} nanowires have been studied recently \cite{SciRep5_7898, SciRep5_14639}. Whereas the effects of magnetic fields on DSMs have been studied previously (for example, in Refs. \onlinecite{EPJB87_92,NatComm5_5161,PRB92_205113} ), these works have tended to focus on magnetic fields perpendicular to the plane of DSM slabs and thin films. In this work, we study the effects of in-plane magnetic fields on DSM thin films. To lay the foundations for our subsequent discussion, we first review the energy dispersion of bulk \ce{Na3Bi} and the emergence of Fermi arcs on semi-infinite slabs of \ce{Na3Bi} terminated perpendicular to the $k$-space separation between the Weyl nodes. We then consider DSM thin films with finite thickness, and next move on to discuss the effects of in-plane magnetic fields on the dispersion relations. We finally showcase one consequence of the magnetic fields by considering the transmission from a source DSM thin film segment without a magnetic field, to a drain DSM segment with a magnetic field for various directions of magnetic fields and source-drain interfaces.  

\section{ Bulk eigenstates } 

The Hamiltonian for bulk \ce{Na3Bi} reads  
\[ 
	H = \epsilon_0(\vec{k}) + \begin{pmatrix} M(\vec{k}) & A k_+ & 0 & 0 \\
	Ak_- & -M(\vec{k}) & 0 & 0 \\
	0 & 0 & M(\vec{k}) & -Ak_- \\
	0 & 0 & -A k_+ & - M(\vec{k}) \end{pmatrix}
\]
where $\epsilon_0 = C_0 + C_1 k_z^2 + C_2 k^2$ and $M(\vec{k}) = M_0 - M_1 k_z^2 - M_2 k^2$, $k^2 = k_x^2+k_y^2$ and $k_\pm = k_x \pm ik_y$  \cite{PRB85_195320, PRB88_125427}. This consists of two uncoupled blocks representing the spin up and spin down states, which we can consider separately. We focus on the spin up states. 

Diagonalizing the spin up block yields the eigenvalues 
\begin{equation}
	\epsilon_\pm = \epsilon_0 \pm \sqrt{ (Ak)^2 + M^2}  \label{egEn} 
\end{equation} 
and the (unnormalized) eigenspinors
\begin{equation}
	\begin{pmatrix}  M \pm \sqrt{ (Ak)^2 + M^2} \\ A(k_x-ik_y) \end{pmatrix}. \label{egSp}
\end{equation} 

\begin{figure}[ht!]
\centering
\includegraphics[scale=0.6]{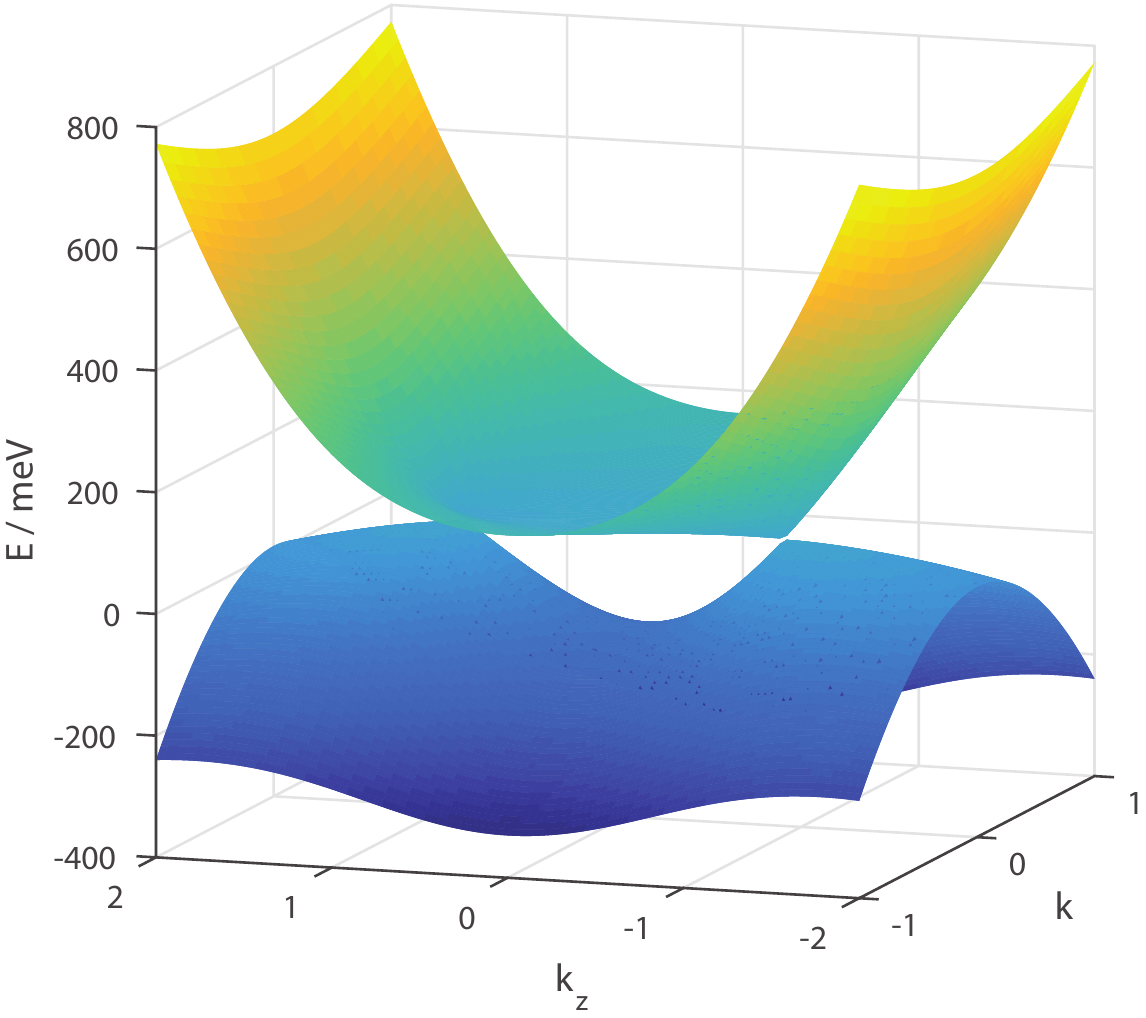}
\caption{  The dispersion relation of bulk \ce{Na3Bi} as a function of $k = \sqrt{k_x^2+k_y^2}$ and $k_z$.  } 
\label{gE4B}
\end{figure}

The bulk energy dispersion is shown in Fig. \ref{gE4B}. The dispersion relation consists of two parabolic cone-like structures above (below) $\epsilon_0$ corresponding to the $+$ ($-$) sign of Eq. \ref{egEn} with the cone tips squashed inwards to form cusps, so that the points where the cone tip starts curving inwards correspond to the lowest (highest) energy at the two Dirac points.  The two Dirac points have energy $\epsilon_0$ and lie along the $k_z=0$ line. 

\section{Semi-infinite slabs}
One hallmark of a DSM is the emergence of Fermi arcs when a bulk DSM is truncated perpendicular to the $k$-space separation between the Weyl nodes. We therefore consider a semi-infinite slab terminated at the $x$ direction so that the slab extends to infinity along the $y$ and $z$ directions, and $k_y$ and $k_z$ are good quantum numbers. 

We look for linear combinations of the eigenstates of the spin up block of Eq. \ref{egSp} which vanish at $x=0$.  
In solving for $k_x$ from the energy equation Eq. \ref{egEn}
 \[
	E_{\pm_{(E)}} = \epsilon_0 \pm_{(E)} \sqrt{ (Ak)^2+M^2 } 
\]
we shift the $\epsilon_0$ to the LHS and square the resulting term, so the $\pm$ distinction in the square root is lost. Here, the subscript $(E)$ indicates that this $\pm$ pertains to $E$ in order to distinguish it from the other $\pm$s that occur later in the text. 

In detail, we have
\begin{eqnarray*}
	&&E = \epsilon_0 \pm \sqrt{ (Ak)^2+M^2 } \\
	&\Rightarrow& (E-\epsilon_0)^2 = (Ak)^2 + M^2 
\end{eqnarray*}
which is a quadratic equation in $k^2$ so that we have, in turn, two values for $k_x^2$ for a given $k_y$. This quadratic equation is cumbersome and not particularly insightful and will not be presented here. We then obtain two values for $k_x^2$, which we designate as $k^2_{x;\pm_{(k_x^2)}}$. In general, only one of these roots will actually be the solution for a given $+/-$ sign value of the $\pm_{(E)}$. 

Now seeking a linear superposition of eigenstates which disappear at the slab surface boundary, we examine the eigenspinors Eq. \ref{egSp}. Within the same $\pm_{(k_x^2)}$ branch, we have two values of $k_x = \pm_{(k_x)} \sqrt{k^2_{x;\pm_{(k_x^2)}}}$. If $k_x$ were imaginary, then one of these will blow up in the wrong direction. If $k_x$ is real, then the upper component of the eigenspinor has the same value for both signs of $\pm_{(k_x)}$, but the $k_x$ in the lower component has opposite signs so it is impossible to form a linear combination of the two such that the eigenspinor vanishes. We therefore conclude that the linear superposition must consist of different $\pm_{(k_x^2)}$ branches.

Next, we investigate whether various quantities are real, complex or imaginary. $E$, $k_z$ and $k_y$ are given to be real. From $E= \epsilon_0 \pm \sqrt{ (Ak)^2 + m^2}$ we conclude that $\epsilon_0$ must be real because $E$ is real and if $\epsilon_0$ has any imaginary component the imaginary component cannot be simultaneously canceled off by both signs of the square root term. $\epsilon_0$ being real in turn restricts $k_x^2$ to be real, so $k_x$ is either real or imaginary.  $E$ being real also constrains $\sqrt{(Ak)^2+M^2}$ to be real. $k^2$ is real, so $M^2$ must be real as well. Thus the upper component of the eigenspinor is real. Earlier we noted that for the wavefunction to disappear at the boundary the two eigenspinors must come from different $\pm_{(k_x^2)}$ branches.  We now know that the upper components of both branches are real, so the relative weights of the two eigenspinors are real. This in turn forces the $k_x$ to be imaginary, so that the lower components of both eigenspinors are imaginary. (Assume that $k_x$ is real so that the only imaginary part of the lower eigenspinor component comes from the $ik_y$ portion. Then for the imaginary part of the lower eigenspinor component to cancel off we need the coefficients of the lower eigenspinor branches to be equal and opposite. This forces $k_x$ to have the same value in both eigenspinors however, which gives the trivial solution of both eigenspinor components being 0.) 

These insights allow us to suggest a numerical scheme to find the allowed values of $(k_y, k_z)$  for a given value of  $E$ for a $\pm x$ terminated slab. An equation in $k_y$ and $k_z$ can be formed in the following way -- for a given $(k_y, k_z)$ we solve for the two values of $k_x$ corresponding to the $\pm_{(k_x^2)}$ branches with the correct sign of the imaginary part depending on whether the semi-infinite slab is terminated along the $+x$ or $-x$ direction. Denoting these two values of $k_x$ as $k_{x_{\pm_{(k_x^2)}}}$, we back substitute them into Eq. \ref{egEn} to determine which of the $\pm_{(E)}$ branches they correspond to, so that the correct form of the eigenspinors can be obtained. With these two eigenspinors, which we denote as $|\pm_{(k_x)^2}\rangle$, we can then calculate the determinant of the linear equations in the unknown coefficients $c_+$ and $c_-$ in $|+_{(k_x^2)}\rangle c_+ + |-_{(k_x^2)}\rangle c_-$. The allowed values of $(k_y,k_z)$ are then those for which the determinant vanishes.

\begin{figure}[ht!]
\centering
\includegraphics[scale=0.6]{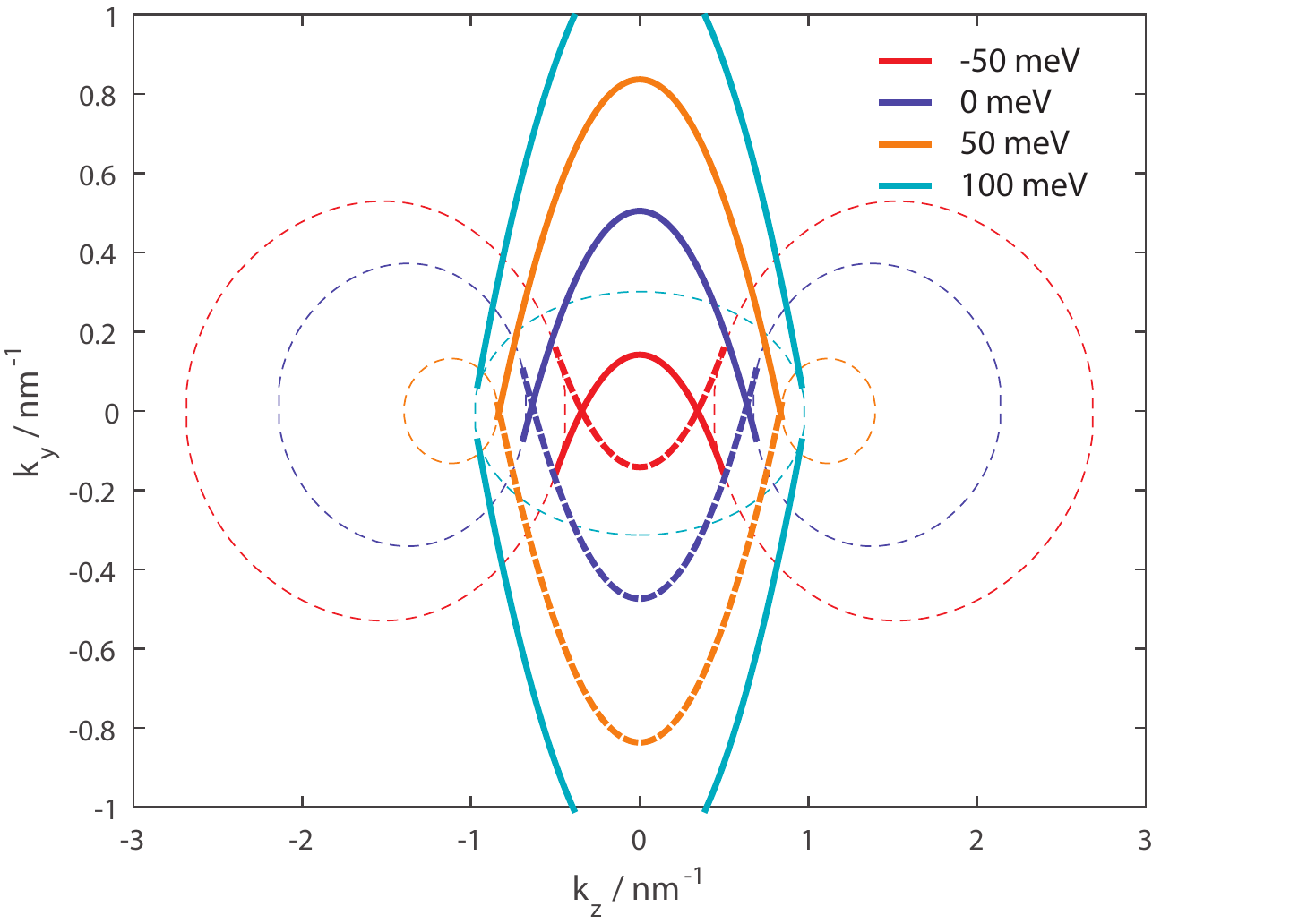}
\caption{  The thick lines indicate the Fermi arcs at the values of energy indicated in the legend. Solid lines correspond to the spin up Fermi arcs at a $+x$ terminated surface on a semi-infinite slab extending to $x\rightarrow -\infty$ while dotted lines correspond to arcs on $x$ terminated surfaces extending to $x \rightarrow +\infty$.  The thin dotted lines are the bulk, infinitely sized blocks at $k_x=0$ at the indicated values of energy.   }
\label{gFermiArcs}
\end{figure}		

Fig. \ref{gFermiArcs} shows the calculated Fermi arcs for spin up at various values of energy for semi-infinite slabs terminated along the $x=0$ line and extending to $x \rightarrow \pm \infty$ respectively, as well as the infnite-sized, bulk \ce{Na3Bi} equal energy contours (EECs) at $k_x=0$ at those values of energy. The spin up Fermi arcs for semi-infinite slabs extending to $x\rightarrow \pm \infty$ are reflections of each other along the $k_y=0$ line. (The spin down Fermi arcs are the reversal of the spin up Fermi arcs) . The arcs emerge once the energy is increased from $-\infty$ to the bottom of the bulk hole band cusp and continue to exist as the energy is increased towards $+\infty$. The arcs are tangential to the bulk EECs calculated with the wavevector perpendicular to the surface ( $k_x$ in this case ) set to 0.  (The two circles, one on the left and the other on the right, for each of the EECs at the three lower values of energies are the cross sections on the two `humps' on either side of the cusp in the hole bands in Fig. \ref{gE4B}. The single circle in the middle for $E=50\ \mathrm{meV}$ is the cross section of the `cone' of the bulk particle band. This value of energy lies above the top of the cusp of the bulk particle band. ) 

We shall later see that the arcs play a significant role in the dispersion relations of thin films.

\section{Thin films} 

We next consider thin films of infinite dimensions along the $y$ and $z$ direction, and finite thickness along the $z$ direction. $k_y$ and $k_z$ are hence good quantum numbers. In contrast to the semi-infinite slab which has only a single surface, a thin film has both an upper and a lower surface. As a result we expect to see aspects which we are familiar with from the previous discussion on semi-infinite slabs, as well as features which now emerge because of the finite thickness. 

We solve the eigenspectrum of the thin film numerically. We adopt the hard-wall boundary conditions under which the wavefunctions vanish at $x=0$ and $x=W$, $W$ being the thickness of the film. Under these boundary conditions, the spatial part of the eigenstate wavefunctions can be expanded as a linear combination of the normalized eigenstates of the infinite potential well $|\phi_n\rangle$ where $\langle x|\phi_n\rangle = \phi_n(x) = \sqrt{2/ W} \sin(n\pi x/W)$.  Since the spin up and spin down states are decoupled in the Hamiltonian, we consider each spin separately.  For each spin and given value of $k_y$ and $k_z$ the relevant part of the Hamiltonian $H(k_y, k_z) _\pm$ (the $\pm$ subscript referring to the spin up / down parts of the Hamiltonian) can be expanded in the basis of the $|\phi_n \rangle$ states. This gives a a numerical matrix $\tilde{H}_{\pm}$ with matrix elements $\tilde{H}_{\pm, [i,j]} = \sqrt{2/W} \int^W_0  \mathrm{d}x \ \sin(i \pi x / W) (H_{\pm} (k_y,k_z) (\sin(j \pi x / W))$. The matrix can then be diagonalized numerically  to obtain the eigenenergies and eigenstates. 

\begin{figure}[ht!]
\centering
\includegraphics[scale=0.4]{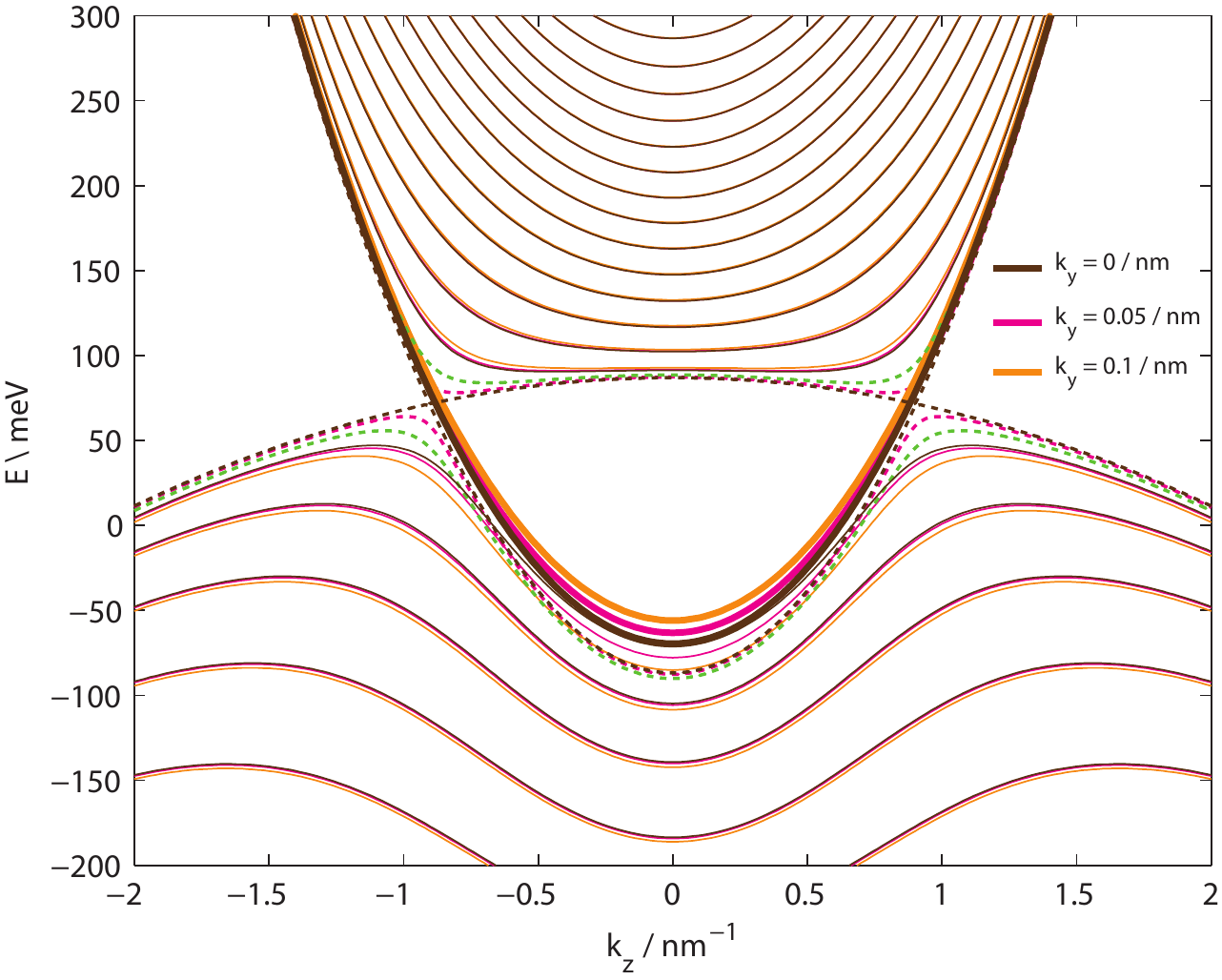}
\caption{  The dispersion relations for a 20 nm thick film as a function of $k_z$ for various values of $k_y$ indicated by the colors of the lines. The dotted lines show the  energy of bulk unbounded \ce{Na3Bi} at $k_x=0$ and the corresponding values of $k_y$ and $k_z$. The thicker line in (a) denotes what we shall, for brevity, refer to as the `lowest energy particle band' (LEPB). } 
\label{gWx20kzComb}
\end{figure}

Fig. \ref{gWx20kzComb} shows the dispersion relations for a 20 nm thick film at various values of $k_y$ as a function of $k_z$.  

Similar to the semi-infinite slab, the hole bands bend downwards in energy at small $|k_z|$ to give rise to a cusp between the two Dirac points.  The finite thickness of the thin film here leads to the formation of subbands. 

There is a subband, highlighted with slightly thicker lines in panel (a) of the figure, which traces the bulk (i.e. infinitely sized DSM block)  $k_x = 0$ particle band for $k_z$ values outside the bulk cusp, and the bulk $k_x=0$ hole band inside the bulk cusp. This band is special, as we shall explain later. For brevity we shall refer to it as the Lowest Energy Particle Band (LEPB), and the subband immediately below it in energy as the Highest Energy Hole Band (HEHB).  Except for the LEPB, the particle band bottoms are bounded in energy above the $k_x=0$ bulk particle band. The energies of the hole bands at $|k_z|$ lying outside the  two bulk Dirac points are also bounded below the $k_x=0$ bulk hole band.  At energies above the Dirac point corresponding to $|k_z|$ lying outside the two bulk Weyl nodes, the energy of the LEPB follows that of the bulk particle states closely. At energies below the Dirac point the energy of the lowest energy particle states lie slightly above the top of the `cusp' in the hole energy bands for $k_z$ lying between the two Weyl nodes. The band bottoms (tops) of the particle (hole) bands increase (decrease) in energy monotonously with $|k_y|$.  

For another perspective we now plot the $k$ space EECs for the 20 nm thick thin film which we have been considering, as well as that of a thicker $50\ \mathrm{nm}$ thick film at $E=-50 \mathrm{meV}$. This is a value in energy which lies above the bottom of the LEPB but below the two `humps' surrounding the small $|k_z|$ cusp.

\begin{figure}[ht!]
\centering
\includegraphics[scale=0.4]{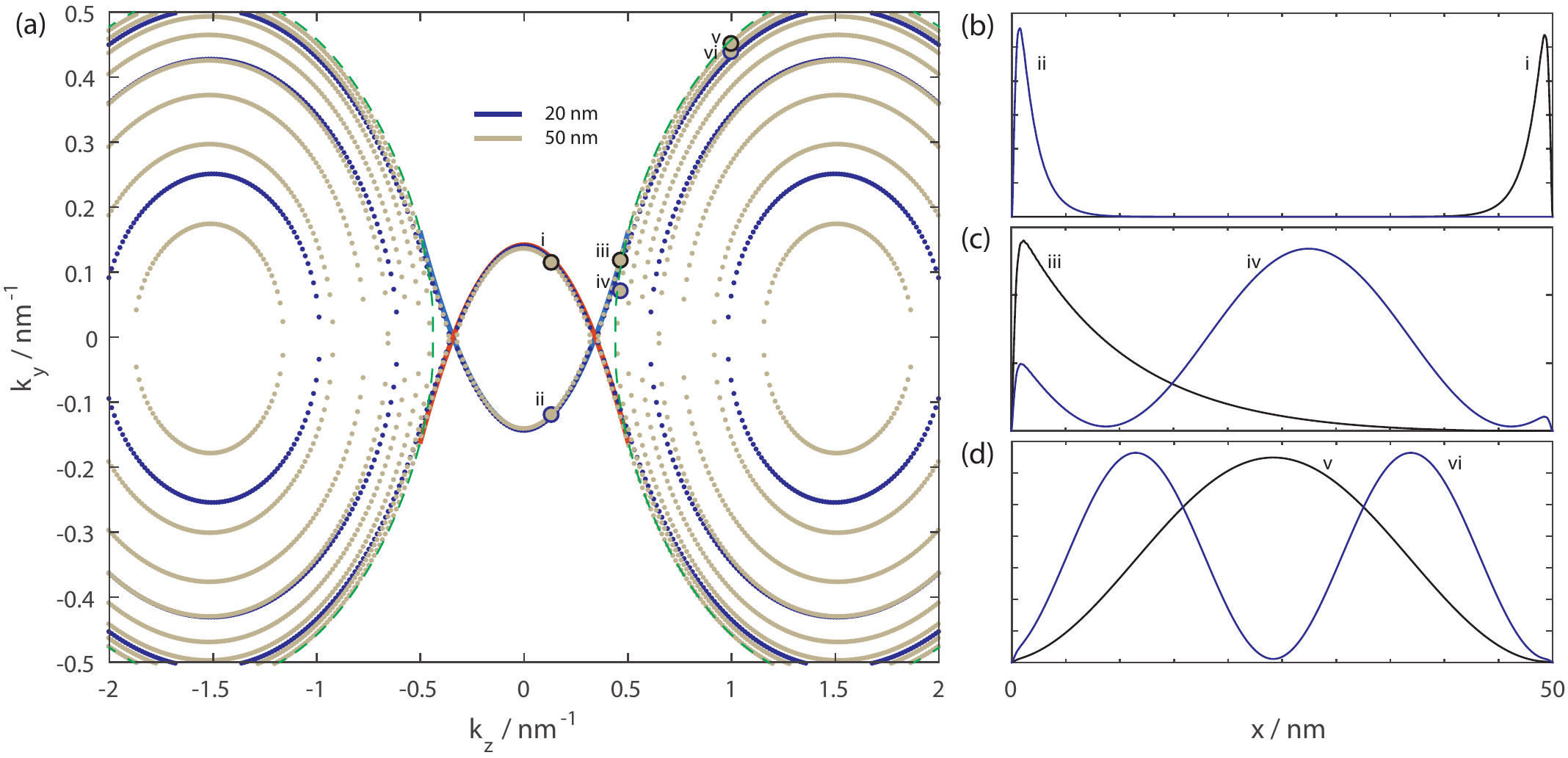}
\caption{  In panel (a), the thick solid lines around the $k$-space origin represent the semi-infinite slab Fermi arcs for slabs infinite in the $y$ and $z$ direction, and semi-infinite in the $+x$ and $-x$ direction at $E=-50 \mathrm{meV}$.  The dotted green lines are the bulk EECs at the same value of energy, while the dots trace out the EECs for a 20 nm, and 50 nm thick film as indicated by the color of the dots. The relative densities of the states at various points in $k$-space and $E=-50 \mathrm{meV}$ marked \textit{i} to \textit{vi} on the EECs are plotted out in the panels on the right. }
\label{gEm50Comb}
\end{figure}		

One dominant feature of EECs is the presence of two series of concentric rings of EECs, one series of rings on the left and the other series on the right which we shall for simplicity refer to as the left and right onions. These two onions correspond to the two `humps' in Fig. \ref{gWx20kzComb} that, at higher energies, eventually form the two Weyl nodes. The empty space between the two onions corresponds to the cusp between the two humps in Fig.  \ref{gWx20kzComb}. The outermost rings of each onion correspond to the HEHB.  The states that lie outside the two semi-infinite slab Fermi arcs are bounded by the bulk EECs indicated by the green dotted lines. These states, such as states (iv) to (vi) in the diagram, are bulk states where there is significant charge density away from the two boundaries. In particular, away from the Fermi arcs, the bulk states (v) and (vi) resemble the usual infinite quantum well states with increasing number of density nodes as we move into the interior of the onion. As one would expect from the infinite quantum well, a larger thickness results in a larger number of states at a given energy. 
   
We now focus on the $k$-space around the vicinity of the semi-infinite slab Fermi arcs. The semi-infinite slab Fermi arcs for the upper and lower surface `penetrate' each other so that for fixed value of $|k_y|<0.15 \mathrm{nm}^{-1}$ a line of constant $k_z$ cuts through both the $+x$ and $-x$ terminated semi-infinite slab Fermi arcs. There are always thin film states which lie near the semi-infinite slab Fermi arcs regardless of the thickness of the film. The states which lie near the Fermi arcs are localized near the surface ( see the states labeled (i) to (iii) ) the arcs correspond to, although the extent of localization decreases with increasing $|k_z|$ (compare state (iii) with states (i) and (ii) ). States (i) and (iii) share approximately the same $k_y$ value. 
 
Point (i) with the smaller $|k_z|$ lies on the LEPB and point (iii) with the bigger $|k_z|$ lies on the HEHB. The states lying in the vicinity of the bulk Fermi arcs for small $|k_z|$ smaller than the $|k_z|$ which the two Fermi arcs intersect each other lie on the LEPB, while the states that lie on the Fermi arcs for larger values of $|k_z|$ lie on the HEHB. The states on the LEPB are therefore always surface states. The states on the HEHB (i.e. the outermost rings of the onions) for small $|k_z|$ lying on the Fermi arcs are also surface states (e.g. iii), but those states on the HEHB with larger values of $|k_z|$ (e.g. vi) are not surface states. 

An examination of the EECs and density distribution at other values of energies (not shown) indicates that the LEPB states at a given energy are always located near the semi-infinite slab Fermi arcs and are localized near either the upper or lower surface, whereas the other states present at the same value of energy located away from the bulk Fermi arcs are bulk states. The $k$-space locations of the LEPB states at a given energy are only very weakly dependent on the film thickness. Together, these indicate that the LEPB states originate from the bulk Fermi arcs and may share the topological protection of the latter. This robustness makes the LEPB states of particular interest in potential device applications. We shall consequently concentrate on the low energy regime in which only the LEPB and a few bulk states are present. 

\section{Effects of an in-plane magnetic field} 
The effects of an out of plane electric field on a DSM thin film / quasi one dimensional nanostructure was investigated in Refs. \cite{SciRep5_7898,SciRep5_14639}. Here we investigate how a magnetic field in the $y$ and $z$ directions affect the dispersion relations. The modification of the EECs by an in-plane magnetic field, and the anisotropy between the $k$-space directions perpendicular and parallel to the $k$ space separation between the Dirac points shall, as we show in the next section, lead to dramatic differences in the transmission from a source DSM segment without a magnetic field to a drain segment with a field as the energy and magnetic field are varied. 

We model a magnetic field in the $y$ direction via the canonical substitution $k_z\rightarrow (k_z +\tilde{B}_y x)$ and a field in the $z$ direction via $k_y \rightarrow (k_y - \tilde{B}_y x)$ where we have absorbed the various physical constants into $\vec{\tilde{B}}$. In this choice of gauge the spatial dependence of the electromagnetic vector potential is in the $x$ direction so $k_y$ and $k_z$ remain good quantum numbers.  In our numerical calculations we adopt units where $e=\hbar=1$, energy is in $\mathrm{meV}$ and lengths in $\mathrm{nm}$.  The units for $k_y$ are in $\mathrm{nm}^{-1}$, so the quantities of $\vec{\tilde{B}}$ which appear subsequently carry units of $\mathrm{nm}^{-2}$.  

\begin{figure}[ht!]
\centering
\includegraphics[scale=0.28]{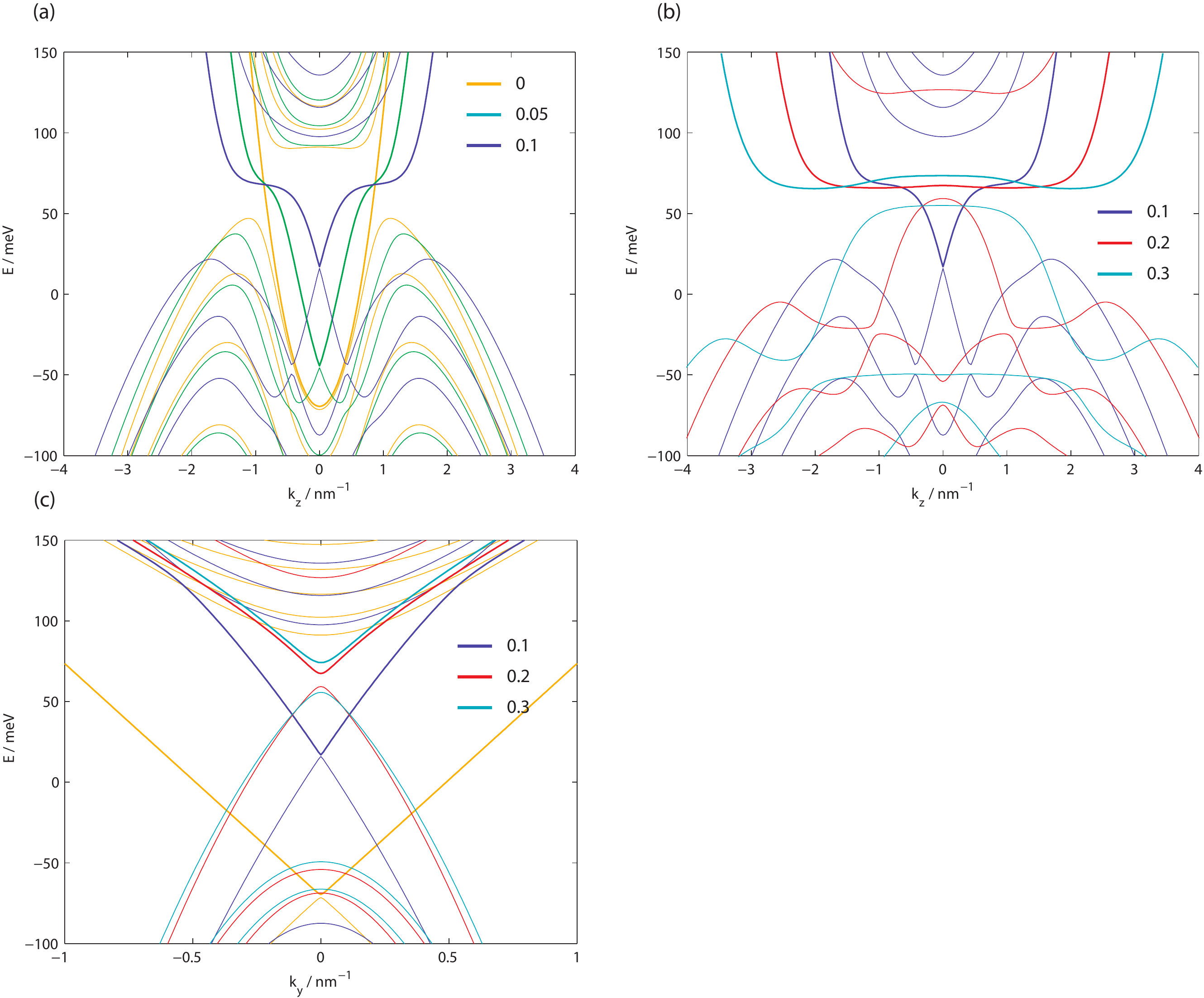}
\caption{  The dispersion relations for a $20\ \mathrm{nm}$ thick slab (a) and (b) as a function of $k_z$ at $k_y = 0$,  and  (c) as a function of $k_y$ at $k_z = 0$ at the values of $\tilde{B}_y$ indicated in the figure legends.  The LEPBs are indicated by thicker lines. }
\label{gW20ByComb}
\end{figure}

Fig. \ref{gW20ByComb} shows the dispersion relations calculated as functions of the $y$ and $z$ wavevectors with the wavevector in the other direction set to 0 at various values of magnetic field in the $y$ direction perpendicular to the $k$-space separation between the two Dirac points. (We have split the $k_y$ dispersion relations into two separate panels for clarity. ) 

Focusing first on panel (a) of the figure, we see that a small $B_y$ leads to the HEHB states with small $|k_z|$ bending upwards in energy to form a small upwards pointing kink centered at $k_z=0$. The band bottom of the LEPB band sharpens to form a downwards pointing kink at $k_z=0$ so that the $k$ space vicinity around where the LEPB and HEHB touch at $\vec{k}=0$ is now reminiscent of a Dirac cone.  The portion of the band on the left and right side of this downwards kink form a plateau at around $60\ \mathrm{meV}$ in energy. The ``Dirac point'' where the LEPB and HEHB touch shifts upwards in energy as $B_y$ increases until it hits the band plateau (see panel (b) ) at around $60\ \mathrm{meV}$.  At this stage the downwards pointing kink in the particle band bottom disappears into the plateau. As $B_y$ increases even further the middle of the LEPB `plateau' becomes concave, the upwards kink of the HEHB blunts out and an energy gap between the LEPB and the HEBB opens up.  Panel (c) of the figure shows that in contrast to the complicated dispersion relation along the $k_z$ direction, the energy of the LEPB increases monotonically with $|k_y|$. We shall show the EECs at various values of energy and $\tilde{B}_y$ in the next section when we discuss the transmission from a source segment without magnetic field to a drain segment with a magnetic field. 

\begin{figure}[ht!]
\centering
\includegraphics[scale=0.28]{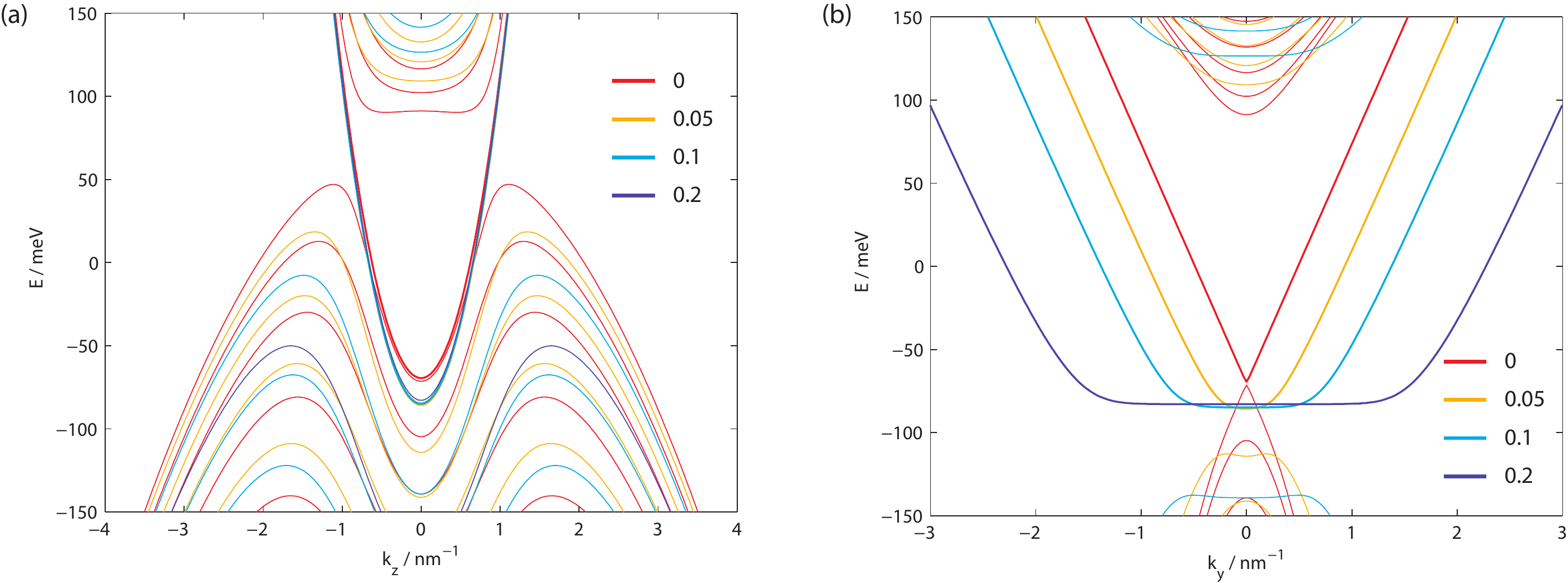}
\caption{  The dispersion relations for a $20\ \mathrm{nm}$ thick slab for (a) as a function of $k_z$ at $k_y = 0$,  and  (b) as a function of $k_y$ at $k_z = 0$ at the values of $\tilde{B}_z$ indicated in the figure legends.  The LEPBs are indicated by thicker lines. }
\label{gW20BzComb}
\end{figure}

Fig. \ref{gW20BzComb} shows the dispersion relations when a magnetic field in the $z$ direction parallel to the $k$ space separation between the two Dirac nodes is applied. 

Panel (a) shows that the dispersion relation of the LEPB along the $k_z$ direction is not very much affected by the magnetic field. A larger field however does push the other particle bands upwards in energy, and the hole bands downwards in energy, as well as increases the energy separation between the subbands. The $z$ magnetic field has a larger effect on the dispersion relation along the $k_y$ direction. As the magnetic field is increased, the band bottom of the LEPB flattens out and forms a plateau for an increasing range of $|k_y|$.  

\section{Transmission} 
\begin{figure}[ht!]
\centering
\includegraphics[scale=0.35]{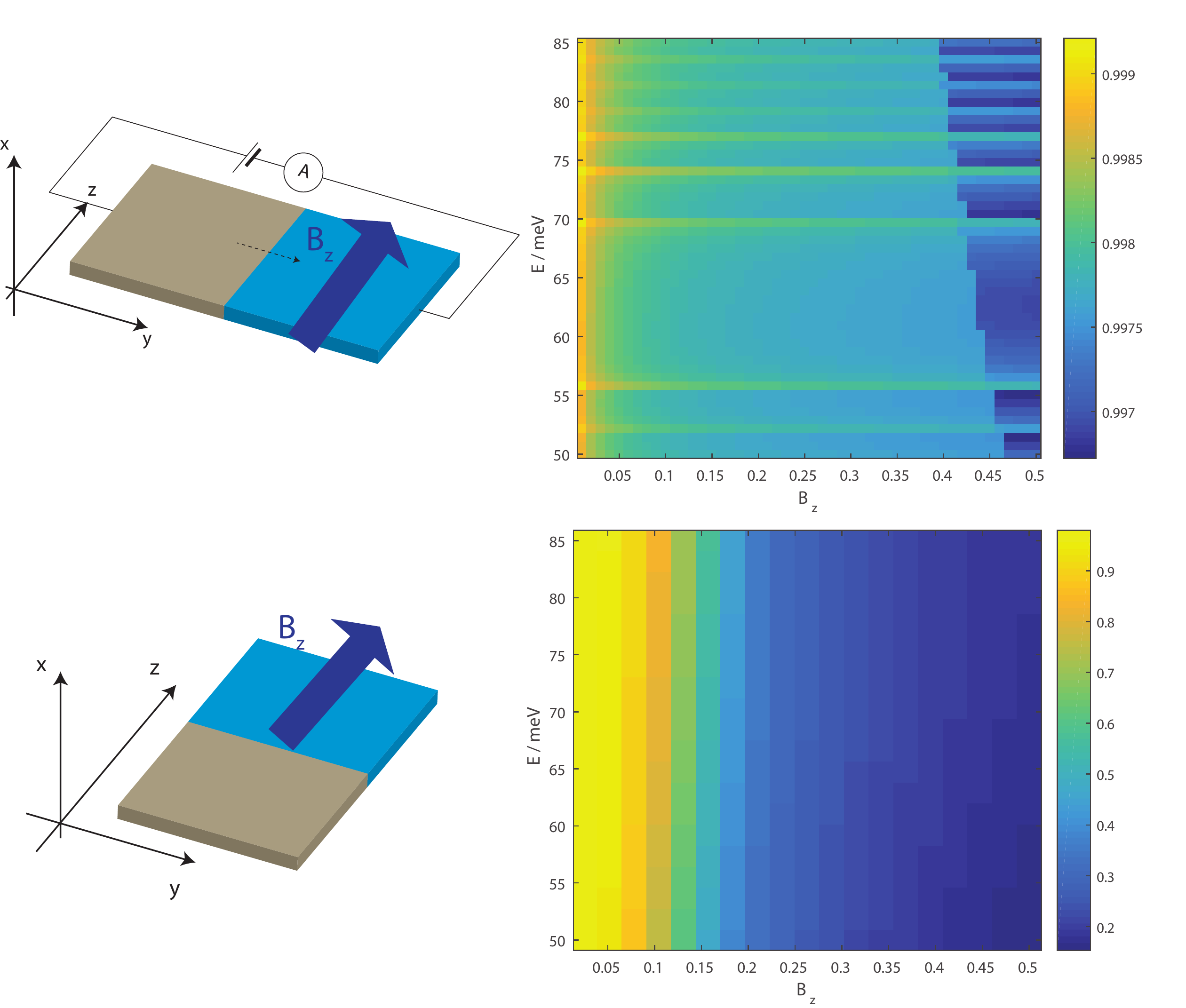}
\caption{  The transmission from a source segment of $20\ \mathrm{nm}$ thick DSM thin film with no magnetic field to a drain segment with a magnetic field in the $z$ direction for (top) an interface parallel to the $z$ direction and (bottom) an interface parallel to the $y$ direction.   }
\label{gTBz}
\end{figure}

Here we consider the transmission of the LEPB states from a source DSM film segment without a magnetic field, to a drain segment with a magnetic field. The setup is shown schematically in the top panel of Fig. \ref{gTBz} where an applied bias, represented by the battery in the figure, drives a current flowing from the source segment to the drain segment. The bias supplies charge carriers at the source segment. The source states incident on the source-drain interface are either transmitted into the drain segment, or reflected back to the source segment. The states transmitted through the interface to the drain result to a current flowing through the circuit, which we represent by the ammeter in the figure. The transmission coefficient for a source state of wavevector $\vec{k}$ and energy $E$ can be calculated by matching the wavefunctions at the source-drain interface. The net transmission at an energy $E$ is obtained by integrating over all the source states of that energy propagating in the direction from the source to the drain.

We consider the cases where the interface between the two segments lies along either the $y=0$ or $z=0$ line, so that the source and drain segments have semi-infinite extents in the in-plane ($yz$) direction perpendicular to the interface and infinite extent parallel to it. The momentum parallel to the interface direction is hence a good quantum number and is conserved in the transmission. Since the integration over all the source states contributing to the transmission at a given energy is essentially an integration over the transverse momentum range spanned by these states, the transmission is very much affected by the overlap between the ranges of transverse momenta spanned by the source and drain states. 

The anisotropy between the $k$-space directions parallel ($z$) and perpendicular to the $k$-space separation leads to different transmission profiles depending on the direction of the interface. 
   
Fig. \ref{gTBz} shows the transmission profiles from a source segment without a magnetic field to a drain segment with a magnetic field in the $z$ direction. (The energy range is chosen so that the only propagating states in the source are the LEPB states. ) The transmission for an interface parallel to the $z$ direction is almost unity and exhibits a decrease with magnetic field for an interface parallel to the $y$ direction.

\begin{figure}[ht!]
\centering
\includegraphics[scale=0.3]{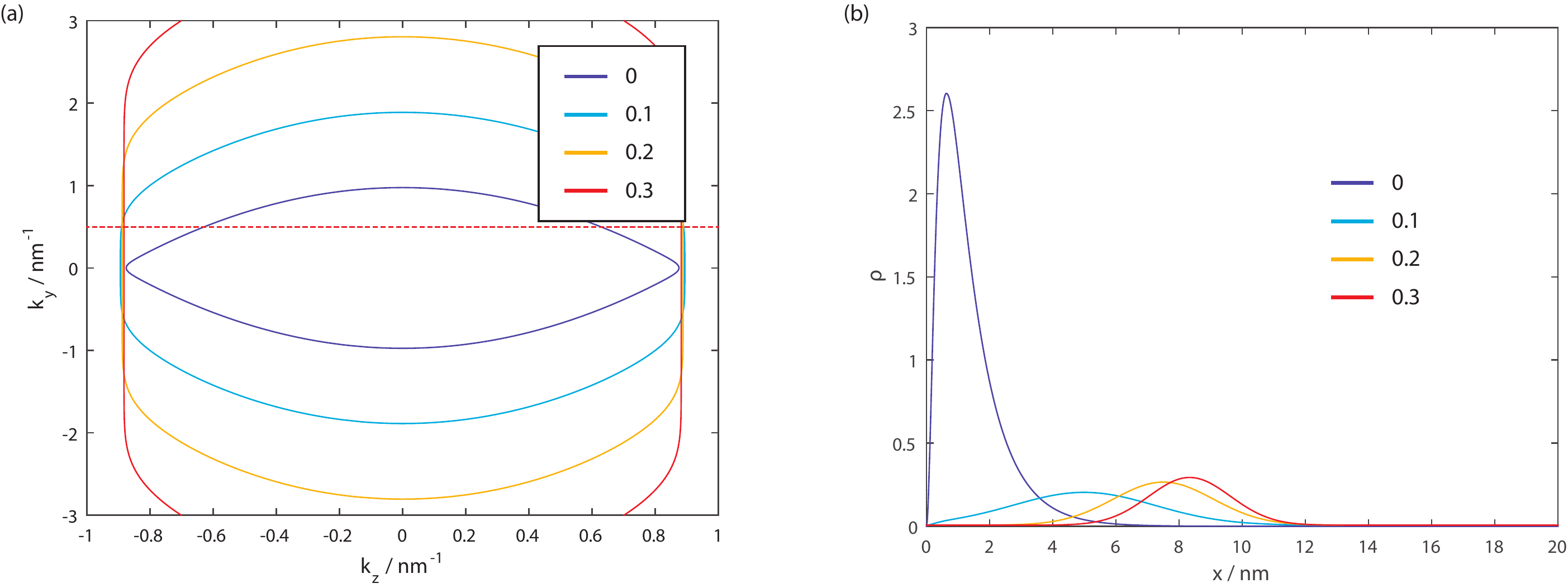}
\caption{  (a) The EECs for a $20\ \mathrm{nm}$ thick DSM film at $70\ \mathrm{meV}$ at various strengths of $B_z$ indicated in the legend. (b) The particle density profiles within the thickness of the thin film at various values of $B_z$ at $k_y = 0.5\ \mathrm{nm}^{-1}$ as indicated by the dotted line on the EEC plots on the (left).   }
\label{gEcEf70BzComb}
\end{figure}

The reason for these trends can be found in the EECs of the source and drain segments shown in Fig. \ref{gEcEf70BzComb}. 

The $k_y$ range spanned by the EECs increases with $B_z$ while the $k_z$ range spanned remains constant. For an interface along the $z$ direction (top of Fig. \ref{gTBz} ) we essentially integrate across the $k_z$ range spanned by the source states in order to calculate the transmission.  At any given value of $k_z$ the profile of the source and drain EECs have a similar trend of curving outwards in the $k_y$ direction. This, and the fact that the $k_z$ range spanned by the source and drain EECs are almost the same and that the EECs have the same shape of bowing outwards along the $k_y$ direction contribute towards the large transmission.  

For an interface along the $y$ direction, we integrate along the $k_y$ range spanned by the source state. Although the $k_y$ range spanned by the drain EECs completely overlaps that spanned by the source EEC, the profile of the EEC at a given $k_y$ between the source and drain segments are quite different -- the source EEC (where $B_z = 0$) curves outwards along the $k_y$ direction but the drain EECs flatten out with increasing $B_z$. This flattening out of the drain EECs is accompanied by a shift in the position along the thickness of the film (in the $x$ direction) where the charge carriers with group velocity from the source to the drain are concentrated. Fig. 8(b) shows that whereas at low fields the charge carriers are localized near the boundaries of the film, they increasingly get shifted towards the interior of the film with increasing $B_z$. This reduces the overlap between the wavefunctions of the LEPB states in the source and drain segments and leads to the trend of decreasing transmission with $B_z$ at a given energy. 

We now shift our focus to the effects of applying a magnetic field in the $y$ direction. 

\begin{figure}[ht!]
\centering
\includegraphics[scale=0.35]{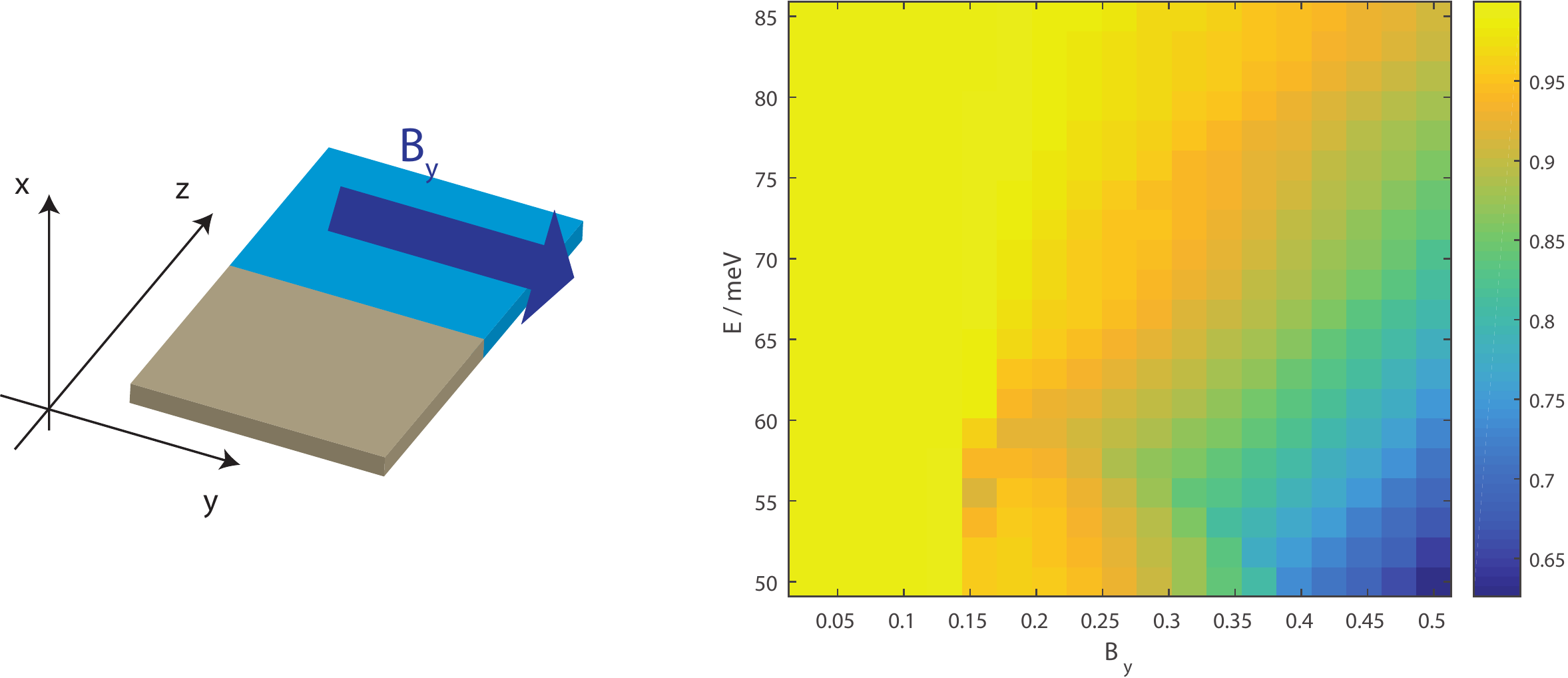}
\caption{  The transmission from a source segment without a magnetic field to a drain segment with a magnetic field in the $y$ direction for an interface parallel to the $y$ direction.    }
\label{gTB101}
\end{figure}

Fig. \ref{gTB101} shows that the transmission from a source segment to a drain segment with a magnetic field in the $y$ direction for an interface parallel to the $y$ direction decreases with increasing magnetic field. 

\begin{figure}[ht!]
\centering
\includegraphics[scale=0.35]{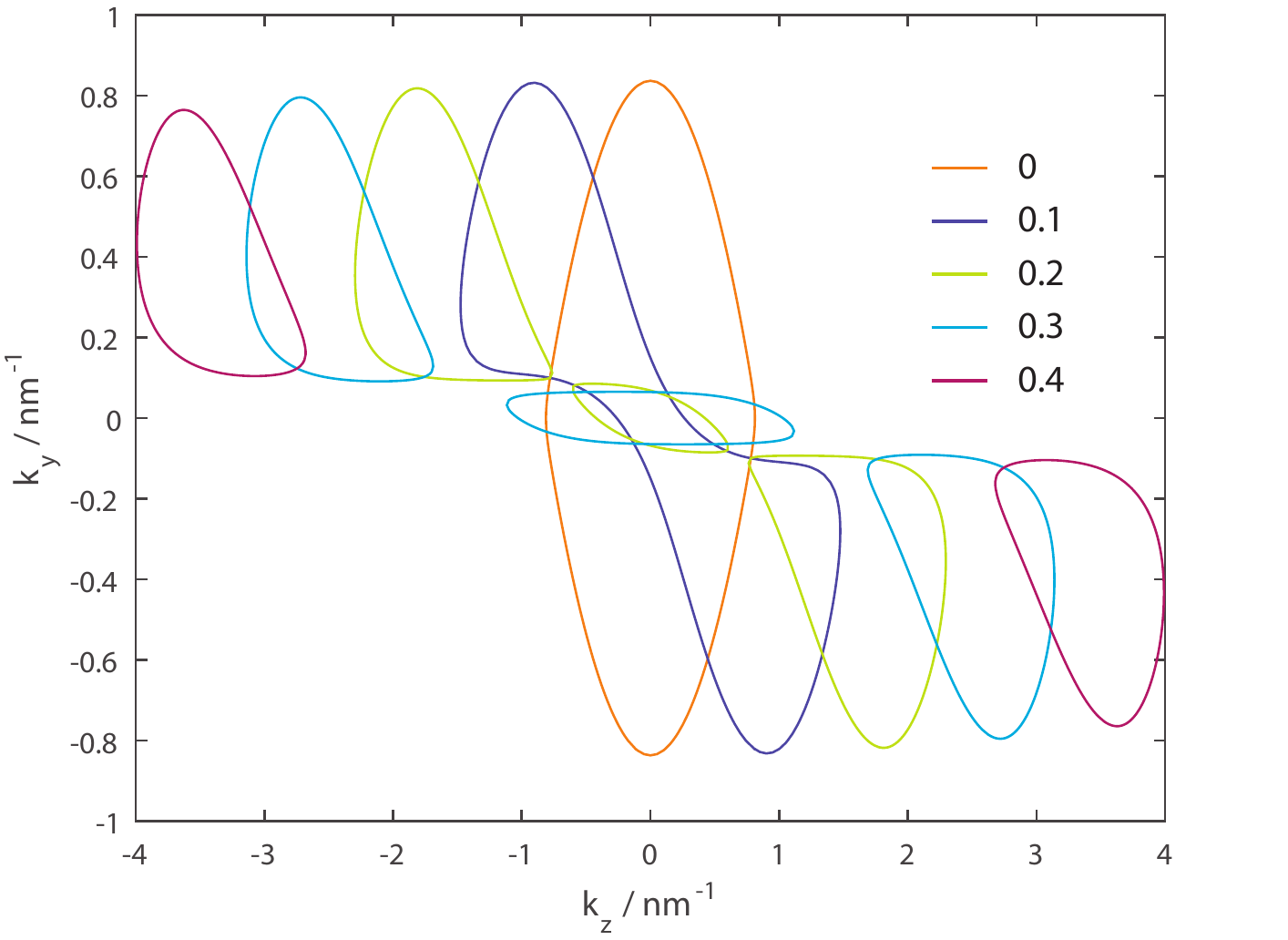}
\caption{  The EECs at $E=50\ \mathrm{meV}$ at various values of $\tilde{B}_z$ indicated in the legend.} 
\label{gEcEf50ByComb}
\end{figure}

This can be explained from the shapes of the EECs as shown in Fig. \ref{gEcEf50ByComb}. 

	Comparing against Fig. \ref{gW20ByComb}, the single lobe for $\tilde{B}_z = 0$ corresponds to the cross section of the LEPB `cone'. The small, narrow `waist' of the $\tilde{B}_z = 0.1$ corresponds to the cross section of the downwards pointing kink in the LEPB. The central lodes enclosing $\vec{k}=0$ surrounded by side lobes for $\tilde{B}_z = 0.2,0.3\ \mathrm{nm}^2$  are the cross sections of the HEHB.  There is no central lobe in the $\tilde{B}_z = 0.4$ because the top of the HEHB has fallen below $E = 50\ \mathrm{meV}$. 
	
	Calculating the transmission across an interface parallel to the $y$ direction involves integrating over the $k_y$ range spanned by the source EEC.  The $k_y$ range spanned by the drain EECs decreases with increasing $B_y$.  As $\tilde{B}_y$ increases from zero, gaps in the EECs begin to open apart so that the EEC which, is originally a single closed curve at $B_y=0$, breaks into three separate curves at $\tilde{B}_y=0.1,0.2\ \mathrm{nm}^2$ in the figure. The gaps in $k$-space between the EEC curves represents an absence of propagating states  at those values of $k_y$ and lead to a drop in the transmission. Once $\tilde{B}_y$ increases beyond around 0.2 where the energy at $\vec{k}=0$ of the LEPB rises above the energy at finite $|\vec{k}|$, the band top of the HEHB drops in energy with increasing $\tilde{B}_y$. An energy gap opens up.  This is manifested as the absence of the central EEC lobe at the larger values of $\tilde{B}_y$ in the figure. At the same time, the $k_y$ ranges spanned by the side lobes decrease with $\tilde{B}_y$. All of these contribute to the drop in transmission with $\tilde{B}_y$.

\begin{figure}[ht!]
\centering
\includegraphics[scale=0.35]{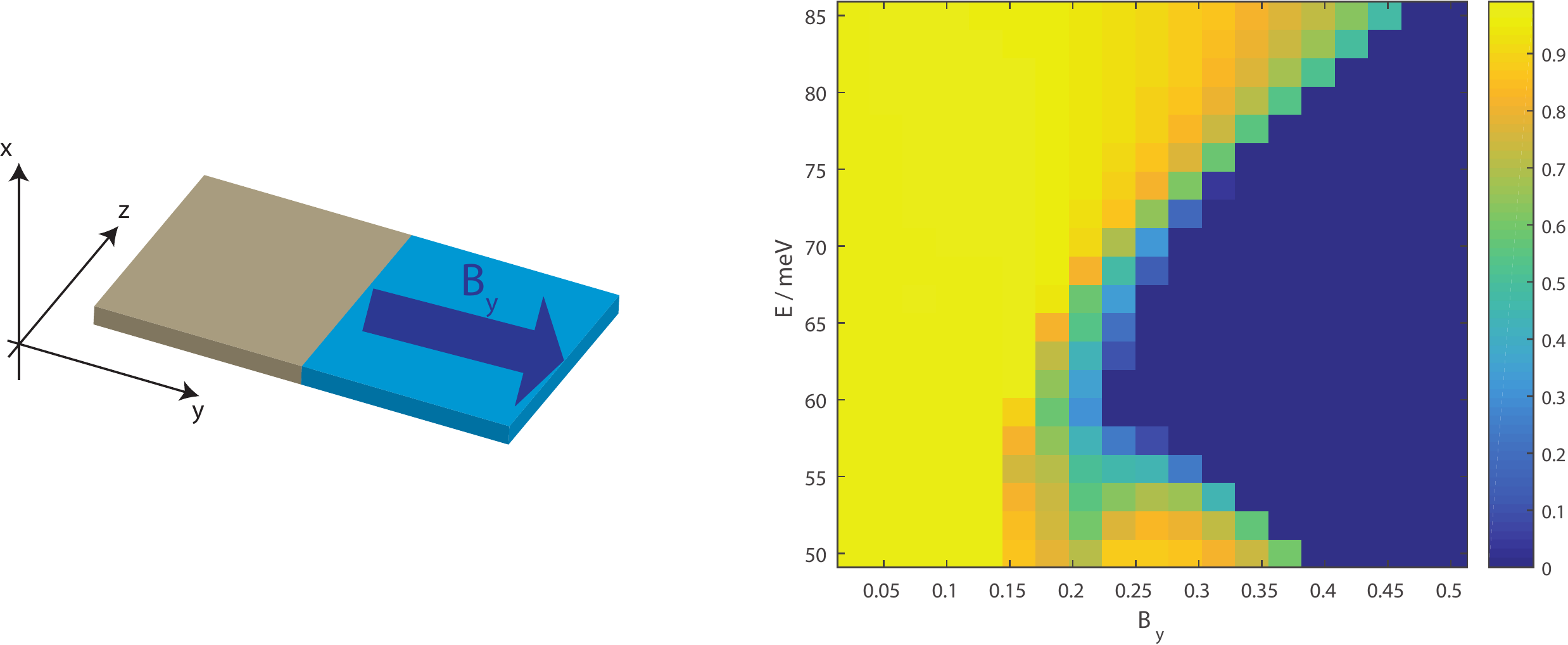}
\caption{  The transmission from a source segment without a magnetic field to a drain segment with a magnetic field in the $y$ direction for an interface parallel to the $z$ direction.    }
\label{gT100}
\end{figure}		

Finally, Fig. \ref{gT100} shows the transmission from a source segment without a magnetic field to a drain segment with a magnetic field in the $y$ direction with the interface parallel to the $z$ direction.  

The decrease in transmission at a given value of energy with $B_y$ can also be explained by the EEC profile in Fig. \ref{gEcEf50ByComb}. Here since the interface is parallel to the $z$ direction $k_z$ is conserved. From the figure it is evident that as $B_y$ is increased the EECs of the drain states are shifted towards larger $|k_z|$ so that the $k_z$ overlap between the source EEC centered around $k_z = 0$ and the drain EECs decreases and eventually drops to 0. 
  
We now look at what happens when we fix $B_y$ and vary the energy. Differing from the transmission profiles considered earlier, here we have at larger values of $B_y$ a trend where the transmission at a given value of $\tilde{B}_y$ first drops to 0 as the energy is increased, and then increases from 0 as the energy is increased further. The region where the transmission is 0 essentially coincides with the energy band gap that opens up between the LEPB and HEHB. 

\begin{figure}[ht!]
\centering
\includegraphics[scale=0.35]{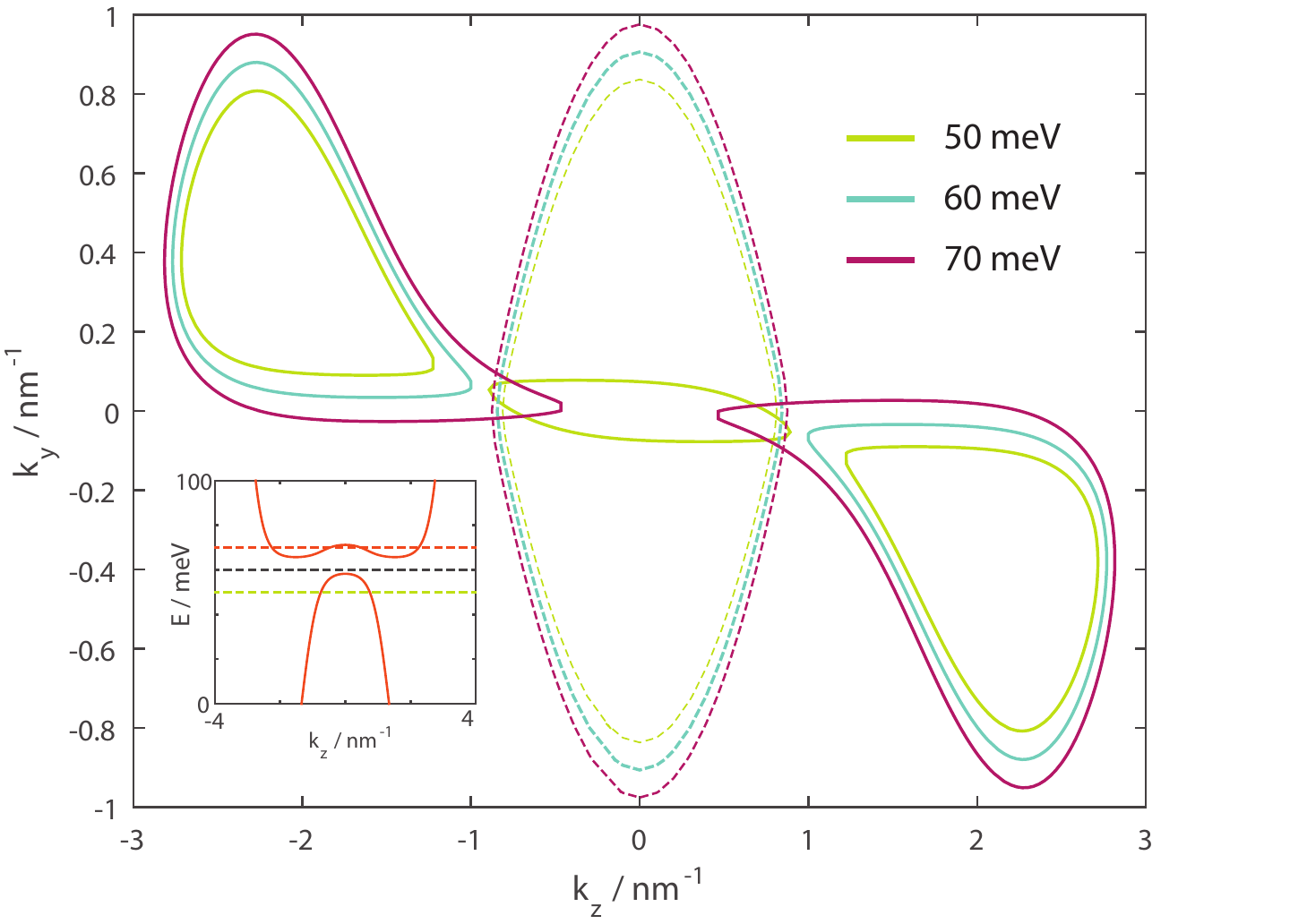}
\caption{  The EECs at $\tilde{B}_y = 0.25$ (solid lines) and $\tilde{B}_y = 0$ at the energies labeled on the legend.  The inset shows the $k_z$ dispersion relation for $\tilde{B}_y = 0.25$ at $k_y = 0$ with the dotted horizontal lines indicating the energy levels at which the EECs are drawn   }
\label{gECBy02comb}
\end{figure}		

This is also reflected in Fig. \ref{gECBy02comb} which show the source and drain EECs at $\tilde{B}_y = 0.025$, and $E = 50, 60$ and $70\ \mathrm{meV}$. These energies correspond to those at the value of $\tilde{B}_y$ where the transmission decreases between $50\ \mathrm{meV}$ to $60\ \mathrm{meV}$, then increases between $60\ \mathrm{meV}$ to $70\ \mathrm{meV}$. The $50\ \mathrm{meV}$ drain EEC has a relative large $k_x$ overlap with EEC due to the middle EEC curve enclosing $|\vec{k}|=0$ which corresponds to the cross section of the HEPB. The next value of energy $60\ \mathrm{meV}$ corresponds to an energy where the $k$-space range enclosed by the source EEC falls in the gap between the HEHB and LEPB and the transmission is zero. The topmost value of energy is above the bottom of the LEPB where the EECs corresponding to the cross section of the LEPB increase in $k$ space area and start to overlap with the source EEC again, giving rise to finite transmission.     

\section{Conclusion} 
In this work we studied the effects of in-plane magnetic fields on a \ce{Na3Bi} DSM thin film. We showed that in the absence of a magnetic field, there is a special band, the LEPB, which originates from the semi-infinite slab Fermi arcs, and is localized near the surface of the thin films. The application of a magnetic field in the $y$ direction leads to the sharpening of the LEPB band bottom and HEHB band top at $\vec{k}=0$, and a shifting of the energy where the LEPB and HEPB meet at $\vec{k}=0$ upwards when the field is first applied. As the field is increased further, the energy of the LEPB band at $\vec{k}=0$ is shifted above that of the $k$-space vicinity and an energy gap opens between the LEPB and HEHB. Applying a magnetic field in the $z$ direction leads to the broadening of the LEPB and the formation of an energy plateau along the $k$ space $y$ direction.  

These changes affect the transmission from a source segment without a magnetic field to a drain segment with a magnetic field as the overlap in the $k$-space direction parallel to the interface between the source and drain EECs varies. In the particular case of an interface parallel to the $y$ direction and magnetic field in the $z$ direction, the $k$ space overlap is not very much affected by the magnitude of the field. The reduction in transmission comes instead from the reduction in spatial overlap between the source and drain wavefunctions when the drain wavefunctions get localized further inside the thickness of the film away from the source wavefunction localized near the film surface.  

After the completion of this manuscript, we learnt of a very recent work \cite{PRB93_081103} on the effects of in-plane magnetic fields on Weyl and Dirac semimetals. In that work, the authors reasoned that an in-plane magnetic field applied parallel to the $k$ space separation between the Weyl nodes (the $z$ direction here) leads to a range of $k_y$ perpendicular to the field in which there is no dispersion along the $k_z$ direction. This agrees with our numerical results in Fig. 8a. Whereas Ref. \onlinecite{PRB93_081103} did not explicitly study an in-plane field magnetic field with only components perpendicular to the Weyl node $k$-space separation, its authors did study an in-plane field applied at an angle to the $k$-space separation. Our results support their conclusion that the components of the in-plane field perpendicular to the $k$-space separation ($B_y$) will lead to a shearing of the EEC along the $k_z$ direction. Our EECs for a magnetic field in the $y$ direction in Fig. 10 shows that $B_z$ pushes the regions of the EEC with differing signs of $k_y$ in opposite directions along the $k_z$ direction. Our results show the additional feature that at large values of $B_z$, this shearing causes the EEC to break up into separate closed curves.   

\section*{Acknowledgment}
The authors acknowledge the Singapore National Research Foundation for support under NRF Award Nos. NRF-CRP9-2011-01 and NRF-CRP12-2013-01, and MOE under Grant No. R263000B10112.


\begin{thebibliography}{99}





	\bibitem{PRB85_195320} Z. Wang \textit{et al}, Phys. Rev. B \textbf{85}, 195320 (2012). 
	\bibitem{PRB88_125427} Z. Wang \textit{et al}, Phys. Rev. B \textbf{88}, 125427 (2013). 

	
	\bibitem{PRB83_205101} X. Wan \textit{et al}, Phys. Rev. B \textbf{83}, 205101 (2011). 
	\bibitem{PRL108_140405} S.M. Young \textit{et al}, Phys. Rev. Lett. \textbf{108}, 14045 (2012). 	

	\bibitem{RMP82_3045} M.Z. Hasan and C.L. Kane, Rev. Mod. Phys. \textbf{82}, 3045 (2010).
	\bibitem{JPSJ82_102001} Y. Ando, J. Phys. Soc. Jpn. \textbf{82}, 102001 (2013). 
	
	\bibitem{NatMat13_677} Z.K. Liu \textit{et al}, Nat. Mater. \textbf{13}, 677 (2014). 
	\bibitem{PRL113_027603} S. Borisenko \textit{et al}, Phys. Rev. Lett. \textbf{13}, 027603 (2014). 
	\bibitem{NatComm5_3786} M. Neupane \textit{et al}, Nat. Commun. \textbf{5}, 3786 (2014). 
	\bibitem{NatMat13_851} S. Jeon \textit{et al}, Nat. Mater. \textbf{13}, 851 (2014). 

	\bibitem{Sci343_864} Z.K. Liu \textit{et al}, Science \textbf{343}, 864 (2014). 
	\bibitem{APL105_031901} Y. Zhang \textit{et al}, Appl. Phys. Lett. \textbf{105}, 031901 (2014). 
	\bibitem{Sci347_294} S-Y  Xu \textit{et al}, Science \textbf{347}, 294 (2014). 


	\bibitem{SciRep5_7898} X. Xiao \textit{et al}, Sci. Rep. \textbf{5}, 7898 (2015).
	\bibitem{SciRep5_14639} H. Pan \textit{et al}, Sci. Rep. \textbf{5}, 14639 (2015).

	\bibitem{EPJB87_92} Philip E.C. Ashby and Jules P. Carbotte, Eur. Phys. J. B 87, 92 (2014).
	\bibitem{NatComm5_5161}  A.C. Potter, I. Kimchi and A. Vishwanath, Nat. Comm. 5, 5161 (2014). 
	\bibitem{PRB92_205113} J. Klier, I.V. Gornyi and A.D. Mirlin, Phys. Rev. B 92, 205113 (2015).
	
	
	\bibitem{PRB93_081103} D. Bulmash and X.-L. Qi, Phys. Rev. B \textbf{93}, 081103 (2016). 	
\end{thebibliography}
\end{document}